# A Group Vehicular Mobility Model for Routing Protocol Analysis in Mobile Ad Hoc Network

Shrirang. Ambaji. Kulkarni and Dr. G Raghavendra Rao

**Abstract**— Performance of routing protocols in mobile ad-hoc networks is greatly affected by the dynamic nature of nodes, route failures, wireless channels with variable bandwidth and scalability issues. A mobility model imitates the real world movement of mobile nodes and is central component to simulation based studies. In this paper we consider mobility nodes which mimic the vehicular motion of nodes like Manhattan mobility model and City Section mobility model. We also propose a new Group Vehicular mobility model that takes the best features of group mobility models like Reference Point Group mobility model and applies it to vehicular models. We analyze the performance of our model known as Group Vehicular mobility model (GVMM) and other vehicular mobility models with various metrics. This analysis provides us with an insight about the impact of mobility models on the performance of routing protocols for ad-hoc networks. The routing protocols are simulated and measured for performance and finally we arrive at the correlation about the impact of mobility models on routing protocols, which are central to the design of mobile ad-hoc networks.

## 1 Introduction

A combination of unique characteristics makes routing in mobile ad-hoc networks challenging. This problem is further compounded when the movements of mobile nodes are restricted to geographic restrictions like pathways and obstacles in mobility models. Mobility models are an integral component of ad-hoc network simulation as they mimic the movement of nodes in real world. Most of the research in the area of mobility characterization has been towards the mobility of individual nodes. As ad-hoc network emphasize a dynamic network among groups, consideration of individual mobility leads to minor significance. Reference Point Group Mobility Model RPGM [1] was one of the successful mobility models which emphasized movements of nodes in groups. It was a model of soldiers following group leader in military scenario. The most popular mobility model used for simulation based studies was the Random Waypoint Mobility Model [2]. In simulation based experiments conducted in [3], [4], it was observed that routing protocols like Ad Hoc On-Demand Distance Vector (AODV), Temporally Ordered Routing Algorithm (TORA) and Dynamic Source Routing (DSR) performed well with mobility models like RPGM. Thus in this paper we extend the RPGM to vehicular motion of nodes restricted by geographic constraints and propose a Group Vehicular Mobility Model (GVMM). We evaluate GVMM with other stochastic mobility models like Manhattan Mobility Model (MHMM) and Freeway Mobility Model (FWMM) with suitable mobility metrics. The impact of proposed mobility model GVMM on routing protocols like AODV, TORA and DSR is also analyzed. The rest of the paper is organized as follows. In Section 2 we discuss of the proposed mobility model GVMM. In Section 3 we analyze the performance of GVMM and other mobility models with suitable metrics. Section 4 simulates the routing protocols

- *Shrirang.Ambaji.Kulkarni is a Research Scholar in the Department of Computer Science and Engineering at National Institute of Engineering Mysore, India.*
- *Dr. G Raghavendra Rao is working as the Principal of NIE Institute of Technology, Mysore, India.*

using GVMM, MHMM and FWMM. Section 5 concludes the paper.

## 2 Group Vehicular Mobility Model (GVMM)

Mobility characterization of a group rather than individual node characterization helps in the design of mobility models. Also most of the routing protocols for ad-hoc networks were developed for group communication and similar applications. Thus we visualize our mobility model to support vehicular movement of nodes as an application supporting a convey of vehicles moving in pathways with various lanes and crossections.

### 2.1 Proposed Algorithm for GVMM

Group Mobility for vehicular movement of nodes requires the following steps.
Algorithm: Group Vehicular Mobility Model
Input: Number of Nodes (N), Number of Groups (G), Speed Deviation and Angle Deviation (A)
1. Read the input parameters like Nodes (N) and number of Groups (G)
2. Read the Speed Deviation and Angle Deviation by which individual nodes deviate from their group leader.
3. Set the movement of the group leader at time t to the motion vector $V^t_{group}$ defined by our vehicular trace file
4. The speed and direction of each group member deviating from its leader is given by

$$Vm(t)=Vl(t)+random() * SDR * max_speed \quad (1)$$

$$\theta m(t)=\theta l(t) + random() * ADR * max_angle. \quad (2)$$

where SDR is speed deviation ratio and ADR is angle deviation ratio
max_speed{10, 20, 30, 40, 50, 60} and max_angle { 0, 2Π } in our experiment
5. Save the movements of nodes with all the details in a trace file for NS-2 simulator.

## 3 Analysis Of GVMM and Other Mobility Models

For our analysis we have considered Manhattan Mobility Model [5] and Freeway Mobility Model [6] along with GVMM.

### 3.1 Mobility Model Metrics

We have considered the following mobility metrics from [6] for our analysis of mobility models.

- Average Link Duration: - This metric specifies the longest interval of time [t1, t2] for nodes i and j forming the link (i, j). This is then averaged for all node pairs for all existing links specifying the equation 3

$$LD' = \frac{\sum_{t1=0}^{T}\sum_{i=1}^{N}\sum_{j=i+1}^{N} LD(i,j,t1)}{P} \quad (3)$$

where P is no of tuples (i, j, t1) and LD (i, j, t1) ≠ 0

- Average Relative Speed:- Relative speed is given by equation 4

$$RST (i, j, t) = |Vi(t) – Vj(t)| \quad (4)$$

where Vi(t) and Vj(t) is the velocity vector of node i and j at time t. The average value of RST(i, j, t) is given by equation 5.

$$RST' = \frac{\sum_{i=1}^{N}\sum_{j=1}^{N}\sum_{t=1}^{T} RST(i,j,t)}{P} \quad (5)$$

where P is no of tuples (i, j, t1) and $RST'$ (i, j, t1) ≠ 0

- Average degree of spatial dependence: - It is a measure of the extent of similarity of velocities of given two nodes not so far apart, given by Ds(i, j, t) and averaged over pair of nodes and time instants and formalized by the equation 6.

$$Ds' = \frac{\sum_{t=1}^{T}\sum_{i=1}^{N}\sum_{j=i+1}^{N} Ds(i,j,t1)}{P} \quad (6)$$

Where P is no of tuples (i, j, t1) and $Ds'$(i, j, t1) ≠ 0

### 3.1 Analysis of Mobility Models

The performance of Group Vehicular Mobility Model, Manhattan Mobility model and Freeway Mobility Model in terms of average link duration is as shown in Figure1.

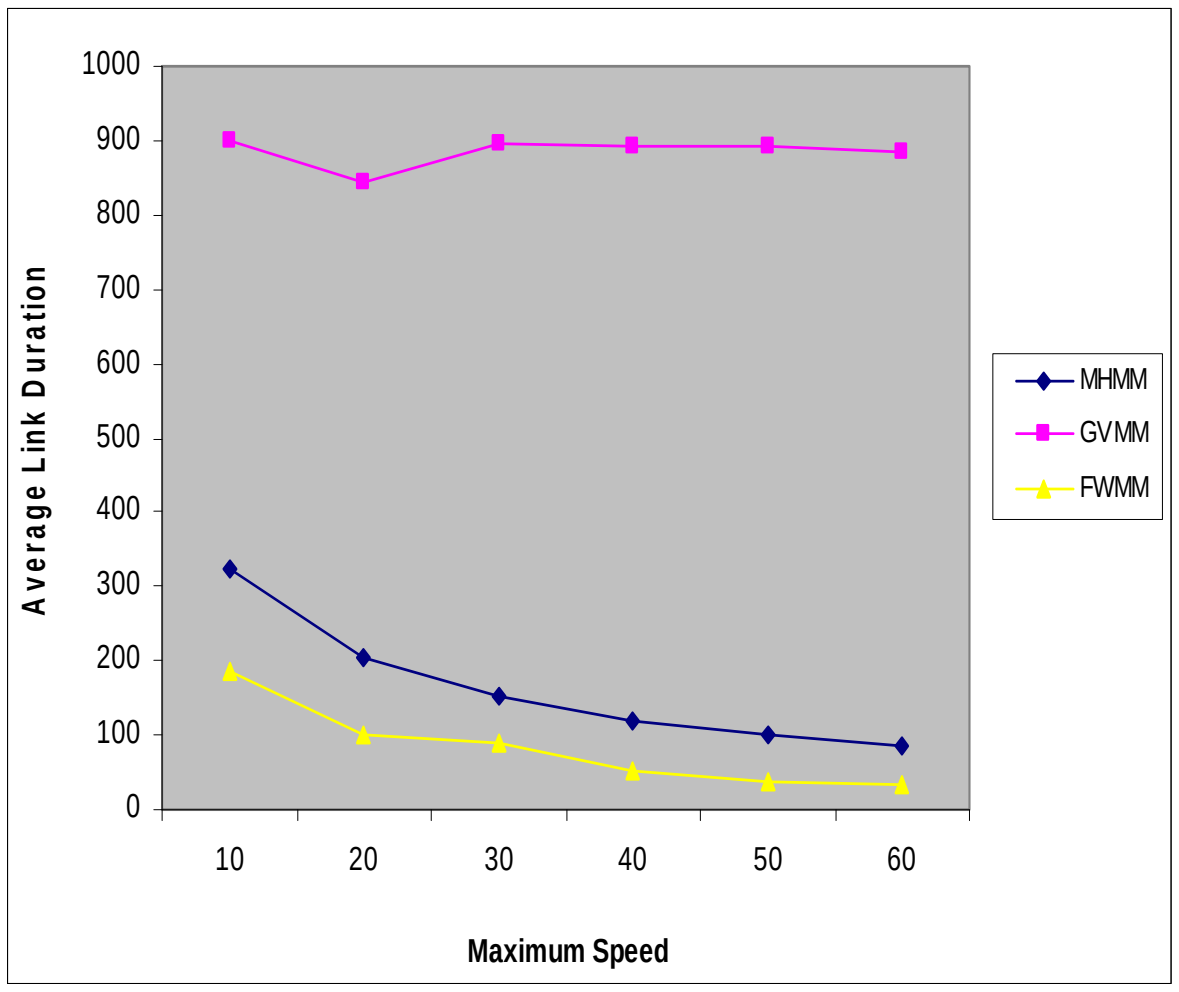

Figure 1: Average Link Duration for Manhattan, Group Vehicular and Freeway Mobility Model.

In Figure 1 we observe that Group Vehicular Mobility Model has a higher value of Average Link Duration as compared to Manhattan or Freeway mobility model. This is because the group of mobile nodes moves at velocities that are deviated from the group leader by small fraction and the existing link between the given two nodes is expected for higher duration.

The average relative speed of the mobility models is as illustrated in Figure 2.

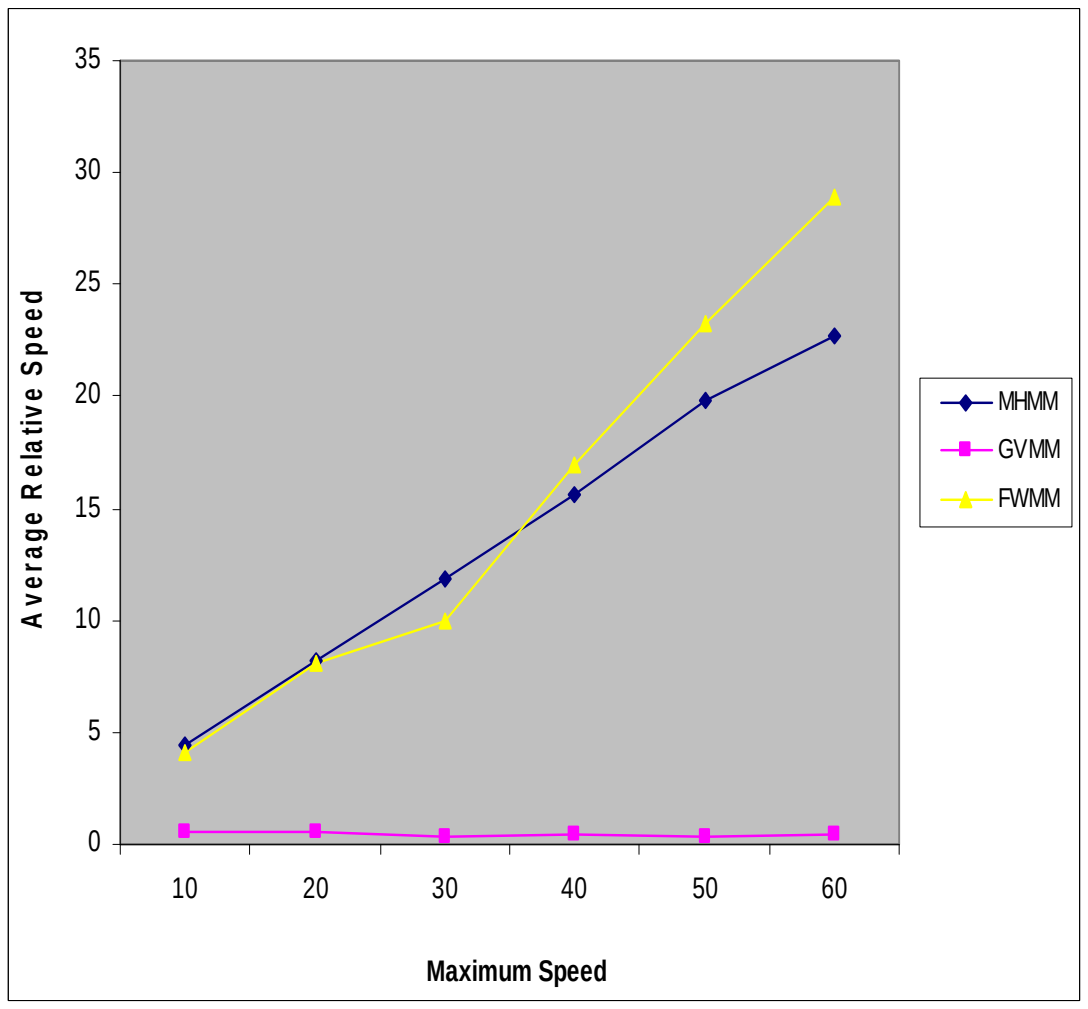

Figure 2. Average Relative Speed for Manhattan, Group Vehicular and Freeway Mobility Model.

In Figure 2 we observe that average relative speed is the lowest for Group Vehicular mobility model as compared to Manhattan and Freeway mobility model. This is because nodes in Group Vehicular move together in a group fashion with minimal deviation among the group nodes. This value is high for both Manhattan and Freeway mobility model as nodes move in opposite direction for both models as lanes exist in opposite direction in their maps.

The average spatial dependency of GVMM, MHMM and FWMM are illustrated in Figure 3.

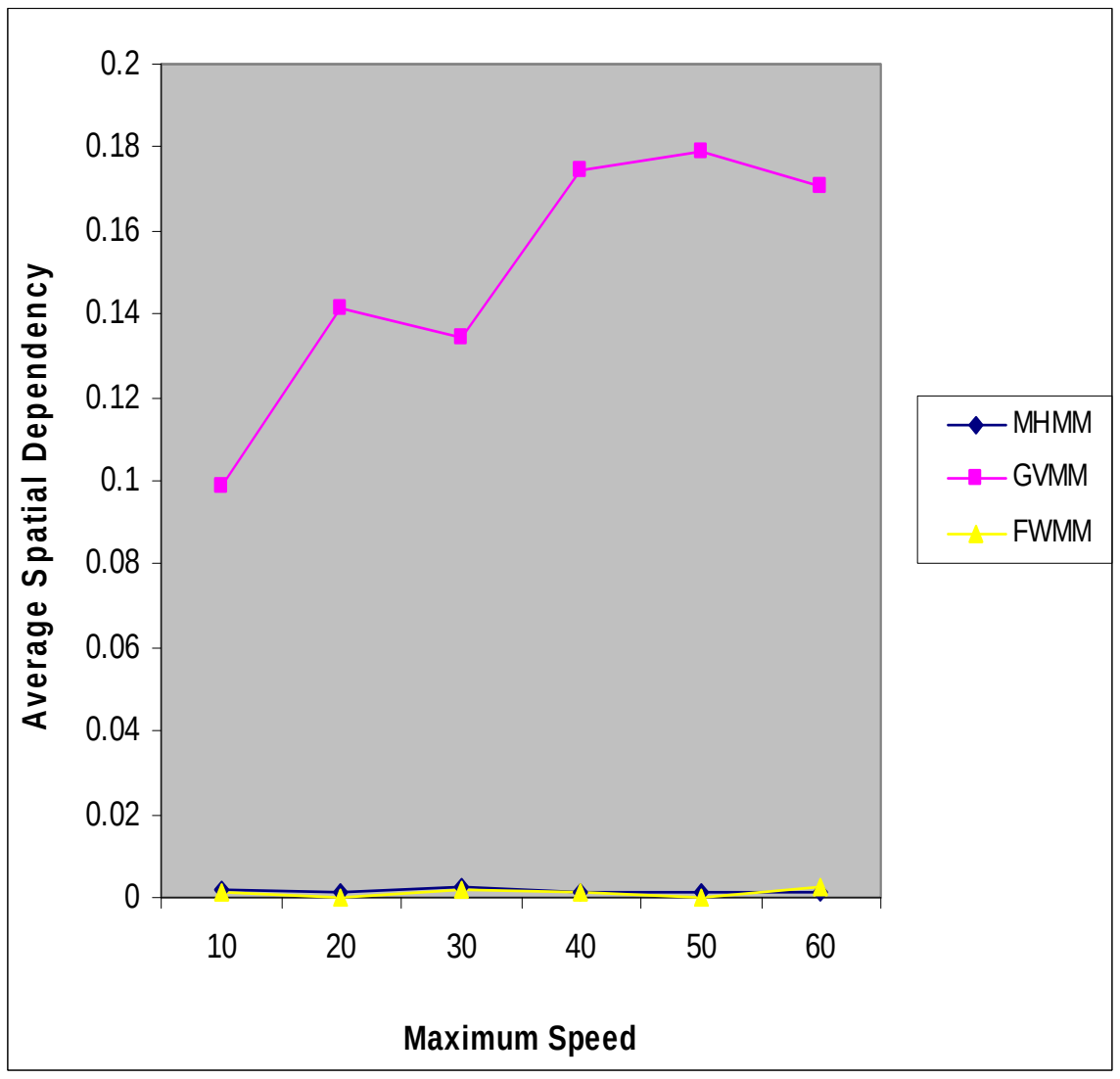

Figure 3. Average spatial dependency for Manhattan, Group Vehicular and Freeway Mobility Model.

In Figure 3 Group Vehicular Mobility model has the highest value for spatial dependency as the group leader controls the movement of mobile nodes even in the presence of opposite lanes whereby nodes in same direction cancel the movement of nodes in the opposite direction. The spatial dependency is very low for both Manhattan and Freeway Mobility model.

Thus we can observe that our proposed mobility model inherits the best performance measures like link duration, relative speed and spatial dependency and helps in complementing the performance of routing protocols as discussed in the next section.

## 4 Simulation of Routing Protocols for Mobile Ad Hoc Networks

Ns-2 simulator ver. 2.29 [7] was used for the analysis of routing protocols like TORA [12], AODV [8], [9] and DSR [10], [11]. The underlying MAC protocol is as defined by IEEE 802.11. Continuous Bit Rate (CBR) traffic sources were used. The mobility models used were Manhattan Mobility Model, Group Vehicular Mobility Model and Freeway Mobility Model. The topology size was 1000 x 1000 m. The traffic generator cbrgen.tcl was applied for the generation of 8 cbr sources at rate of 4.0 kbps. The number of nodes in the simulation was 50. Each mobility model had 6 corresponding scenario file with maximum speed varied from 10 to 60 secs.

### 4.1 Routing Performance Metrics

The metrics used for analysis were derived from [13] for detailed protocol performance analysis.

- Packet Delivery ratio: - The ratio between the number of packets originated by the application layer to those delivered to the final destination.
- Routing overhead (Normalized Routing load):- The number of routing packets transmitted per data packet delivered to the destination.
- Path Optimality( Average End-End Delay):- The difference between the number of hops a packet took to reach its destination and the length of shortest path that physically existed through the network when the packet was originated

Packet delivery ratio is an important metric in terms of the robustness of the routing protocol. Routing overhead or normalized routing load measures the scalability of the protocol. Path optimality is the ability to use the network resources by selecting an optimal path from the source to the destination.

### 4.2 Simulation Results

The performance of AODV, TORA and DSR on Manhattan Mobility Model is as shown in Figure 4 (a), 4(b) and 4(c).

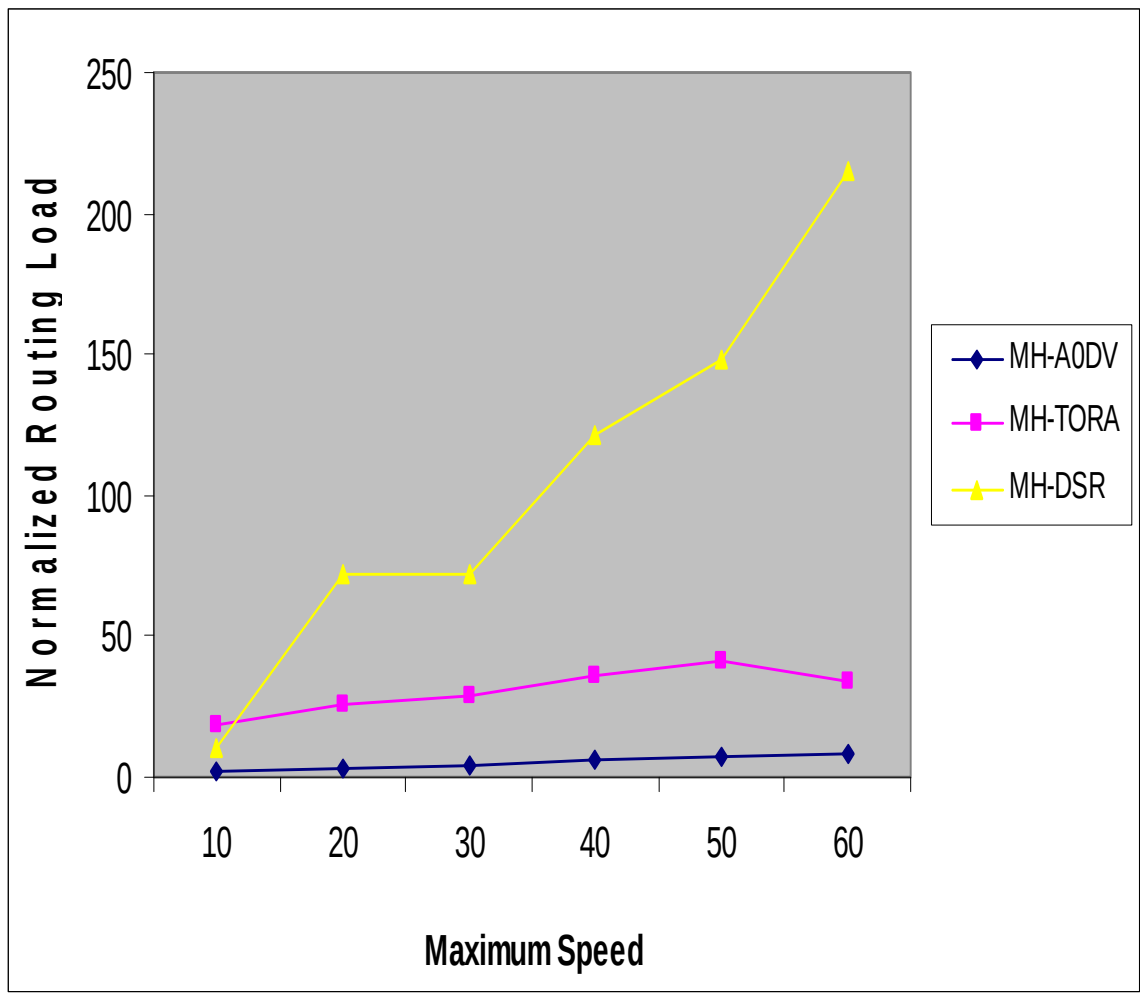

Figure 4(a) Normalized Routing Load for AODV, TORA and DSR on Manhattan Mobility Model

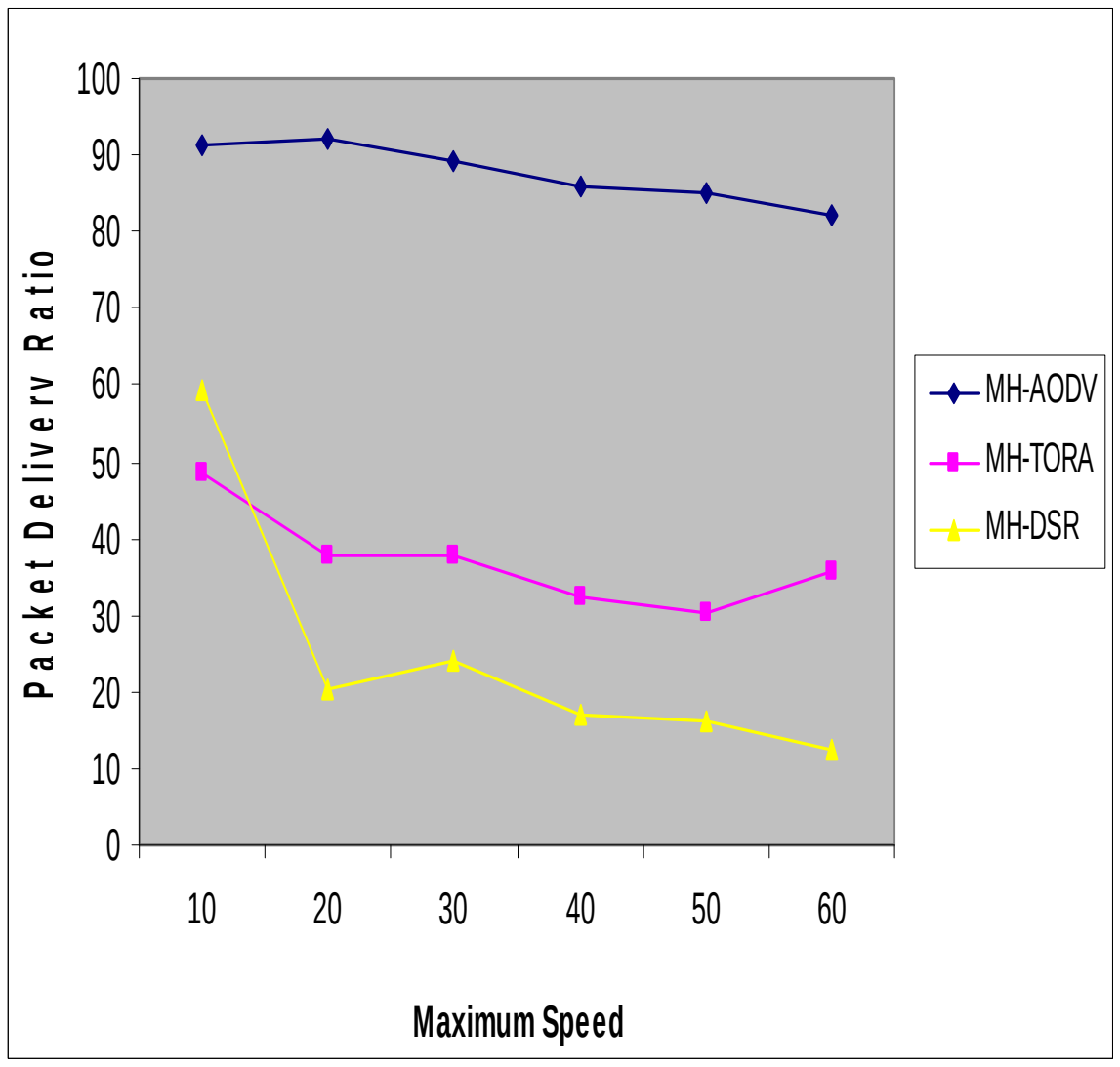

Figure 4(b) Packet Delivery Ratio for AODV, TORA and DSR for Manhattan Mobility Model

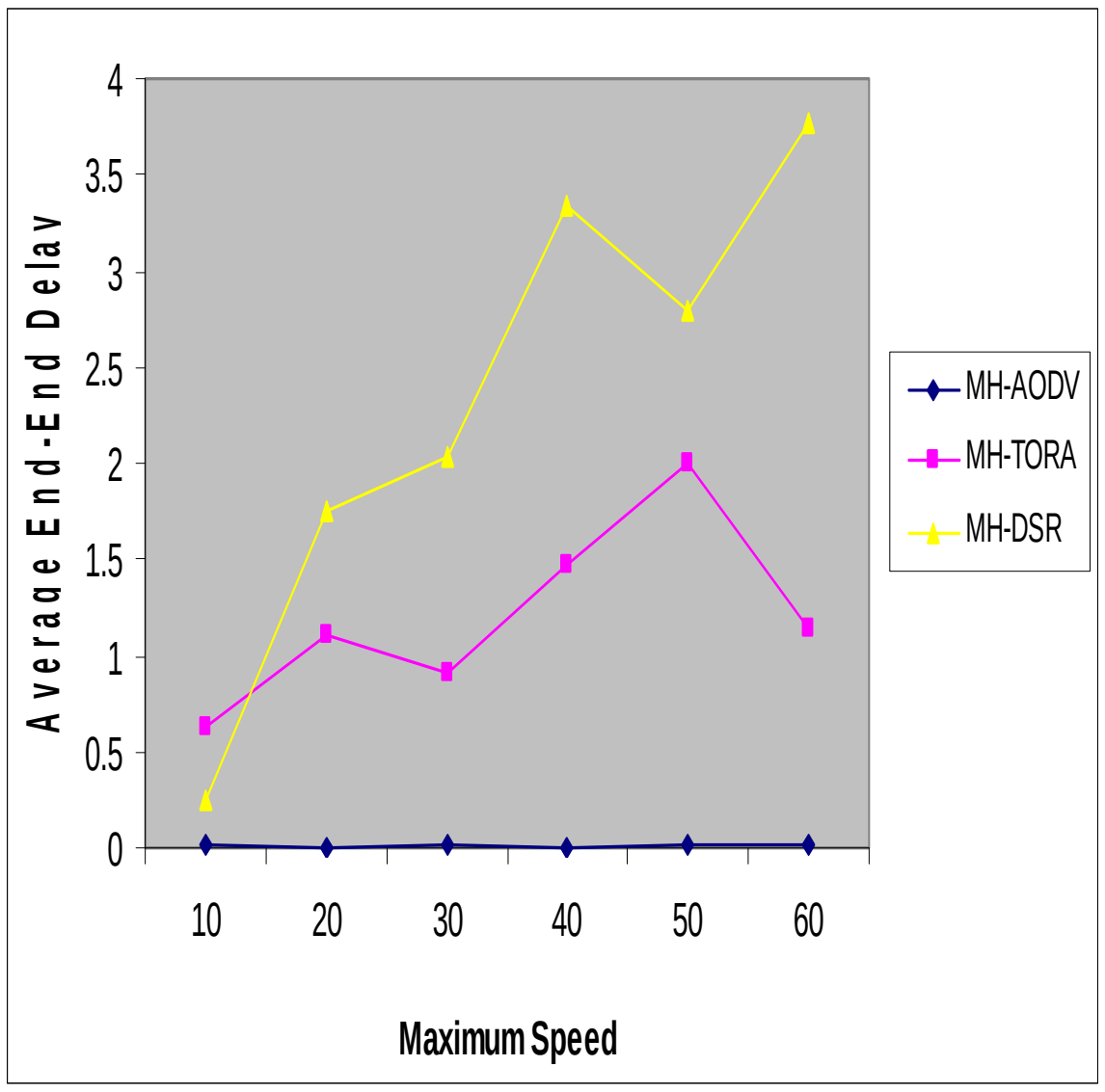

Figure 4(c) Average End-End Delay for AODV, TORA and DSR on Manhattan Mobility Model

In Figure 4(a), 4(b) and 4(c) we observe that AODV outscores all other routing protocols in terms of routing overhead, packet delivery ratio and end-end delay. We observe in 4(a) that at high speed the routing overhead in DSR is high because of route cache becomes stale quickly. Also in 4(c) the route delay is very high for DSR as DSR uses the length of the route as the main criteria for choosing a route from several routes whereas the delay is least for AODV because it prefers the least congested route. The packet delivery ratio in Figure 4(b) indicates the poor performance of DSR at high speed the rate of link failures can happen frequently. Also the high relative speed of Manhattan mobility model means lower link duration and lower throughput and high overhead.

The performance of AODV, TORA and DSR on Group Vehicular Mobility Model (GVMM) is illustrated in Figure 5(a), 5(b) and 5(c).

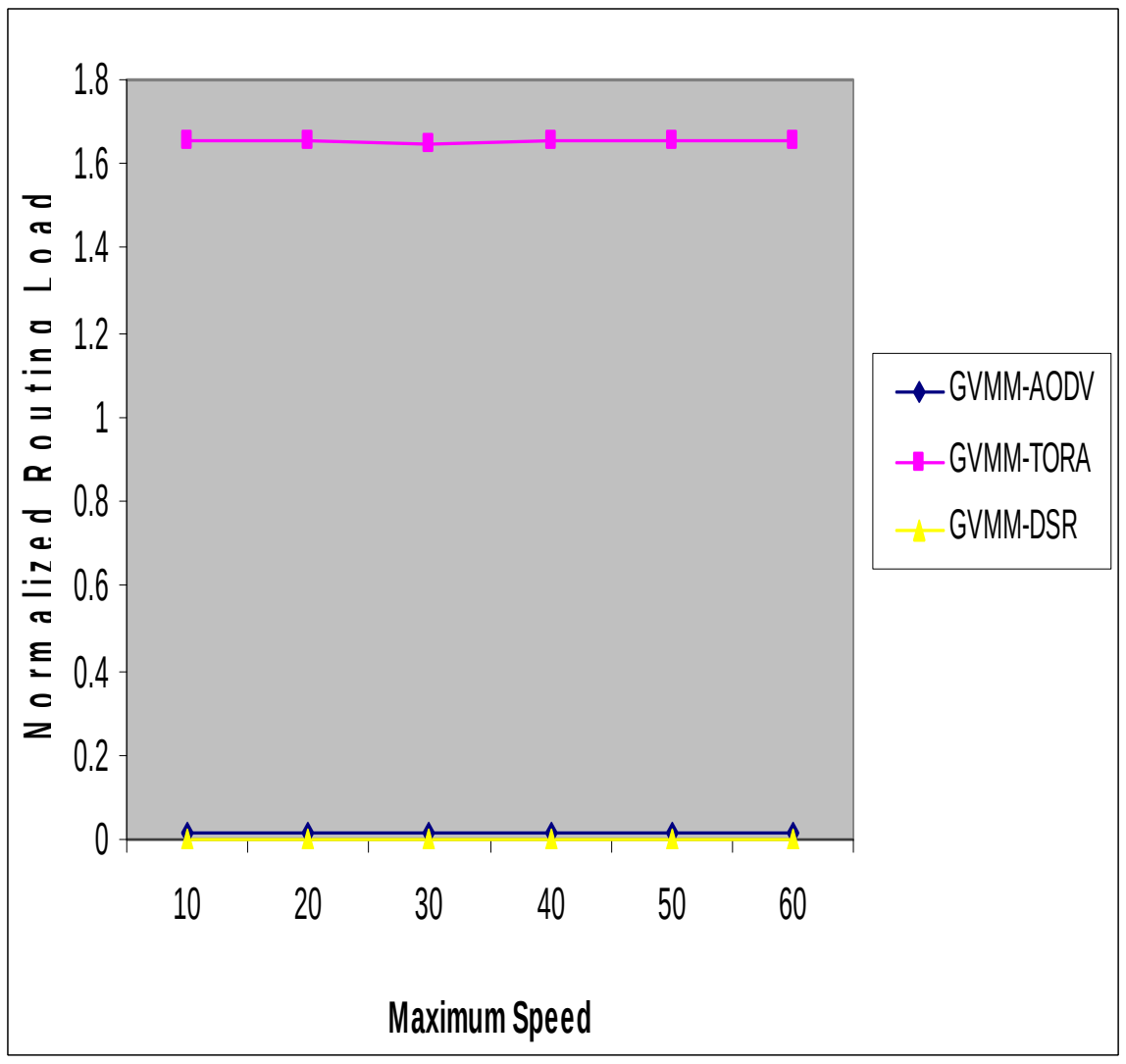

Figure 5(a) Normalized Routing Load for AODV, TORA and DSR on Group Vehicular Mobility Model

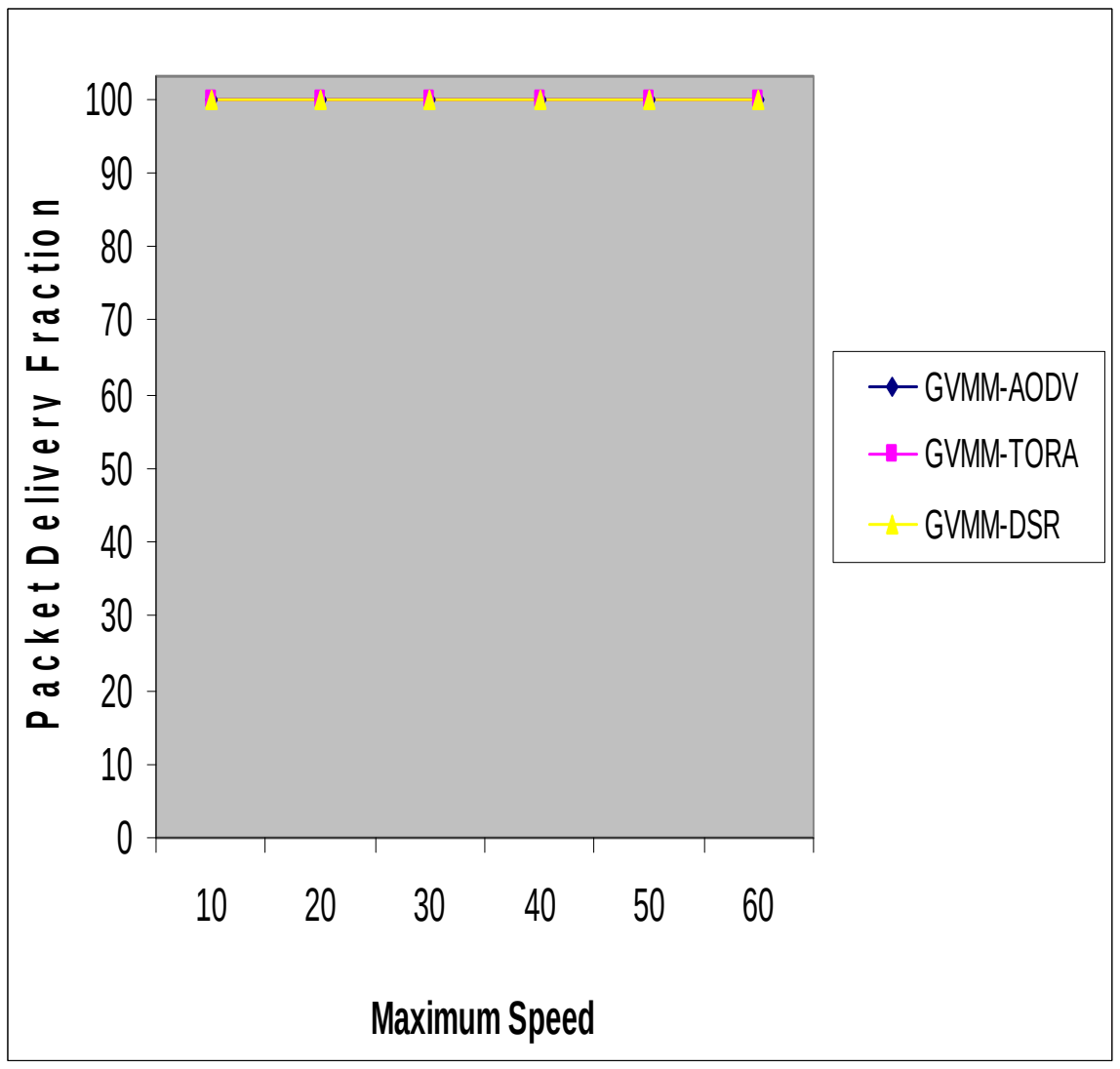

Figure 5(b) Packet Delivery Ratio for AODV, TORA and DSR for Group Vehicular Mobility Model

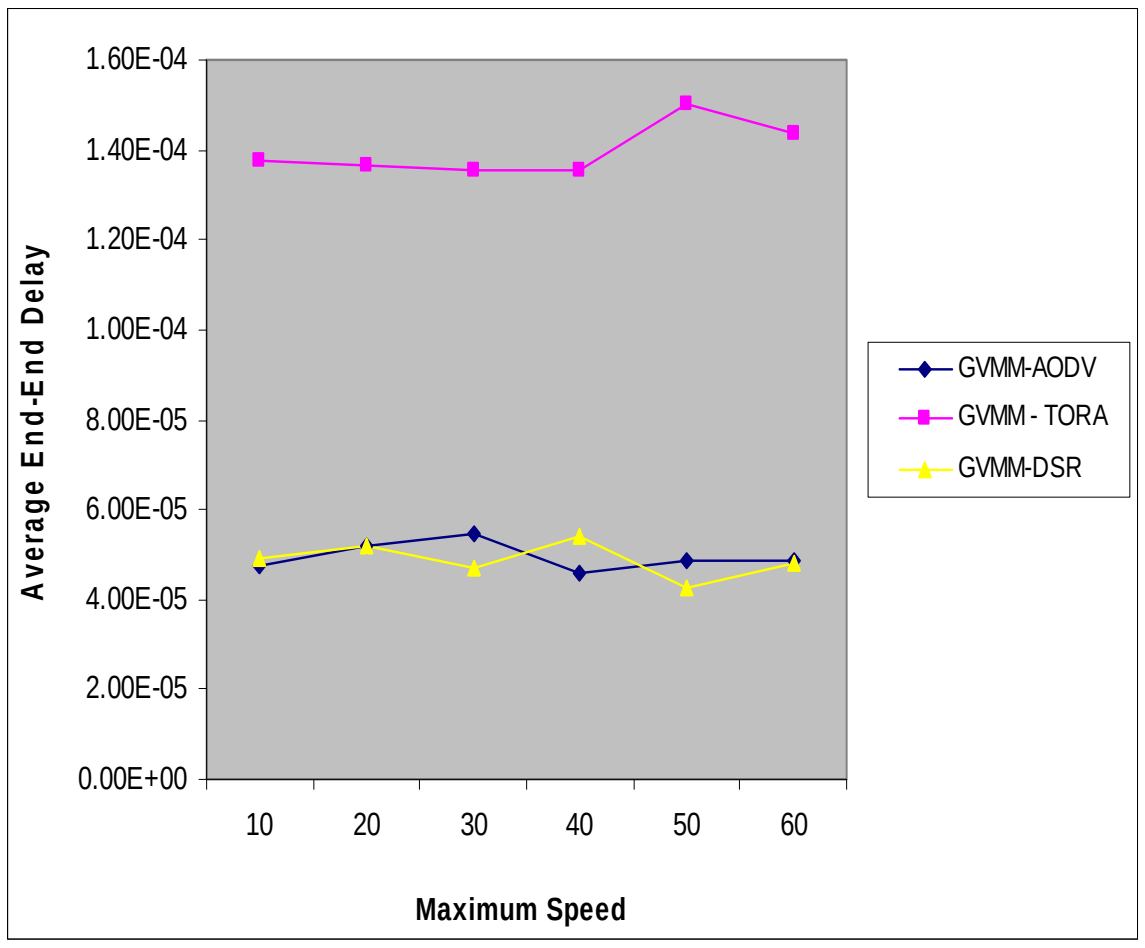

Figure 5(c) Average End-End Delay for AODV, TORA and DSR on Group Vehicular Mobility Model

In Figure 5(a), 5(b) and 5(c) we observe that the performance of TORA in terms of routing load and end-end delay is poor. This is because TORA consumes more overhead for large number of nodes. Regarding path optimality TORA was not designed for path optimality from the outset. The packet delivery ratio of all the three protocols is appreciably high on Group Vehicular Mobility Model. The packet delivery ratio in Figure 5(b) is high because of high spatial dependency which in turn leads to high link duration and the corresponding high packet delivery ratio

The performance of AODV, TORA and DSR on Freeway Mobility Model is as illustrated in Figure 6 (a), 6(b) and 6(c).

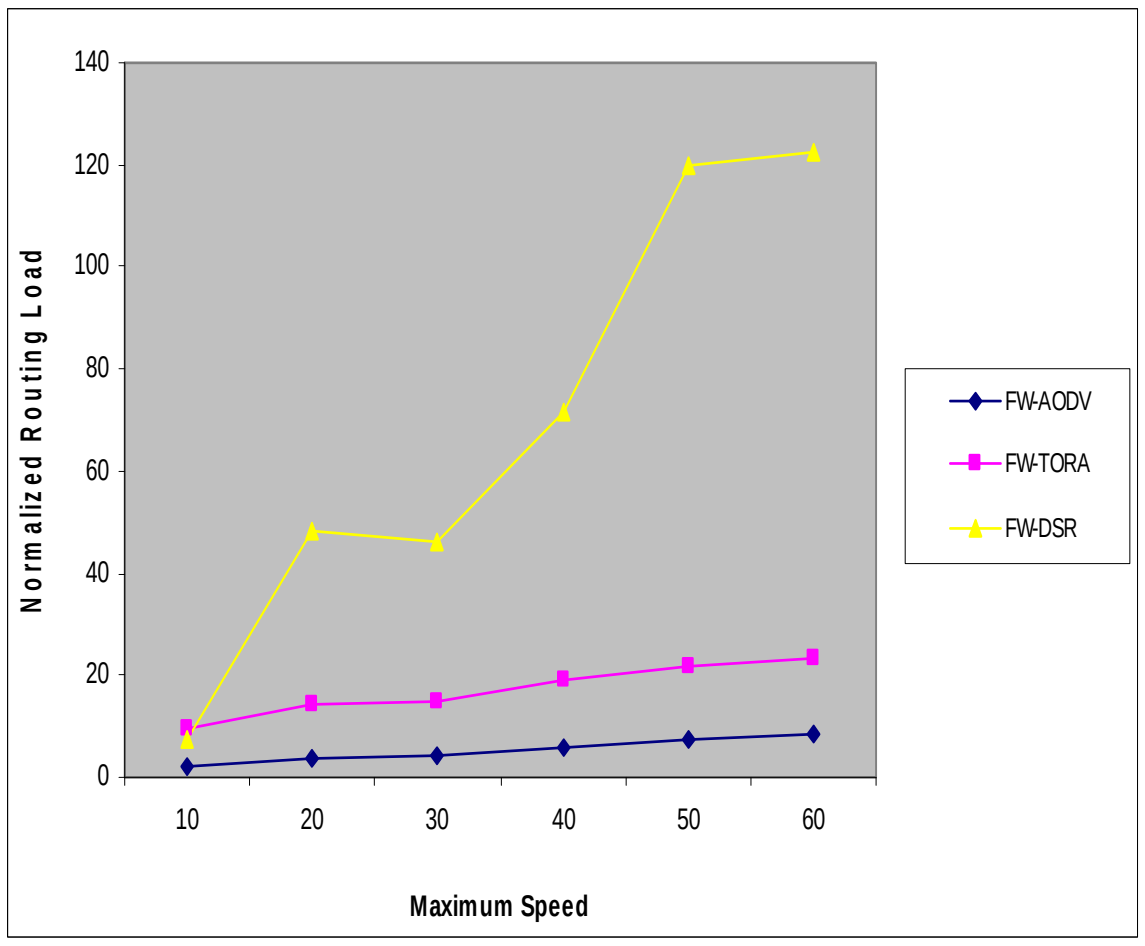

Figure 6(a) Normalized Routing Load for AODV, TORA and DSR on Freeway Mobility Model

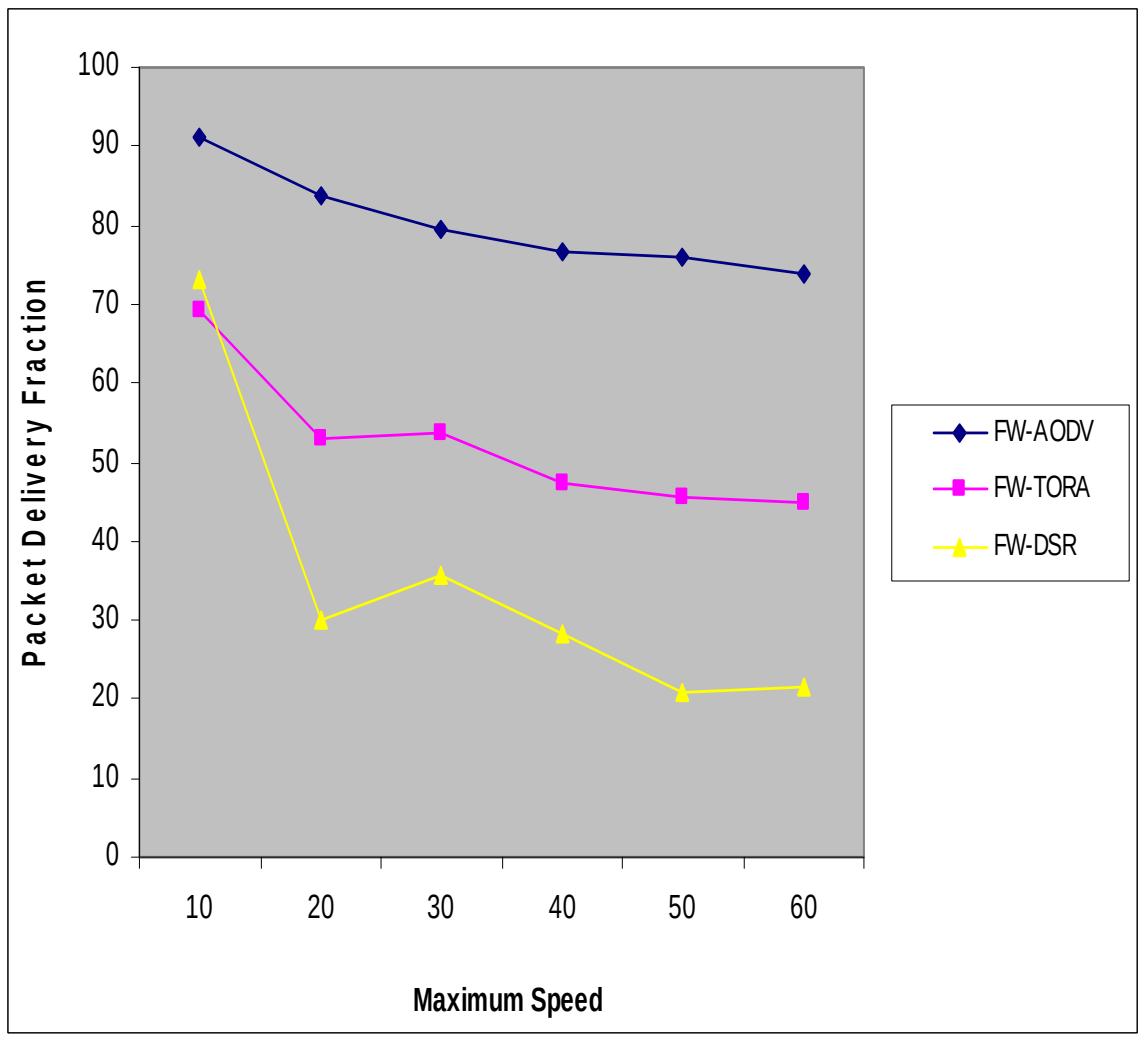

Figure 6(b) Packet Delivery Ratio for AODV, TORA and DSR on Freeway Mobility Model

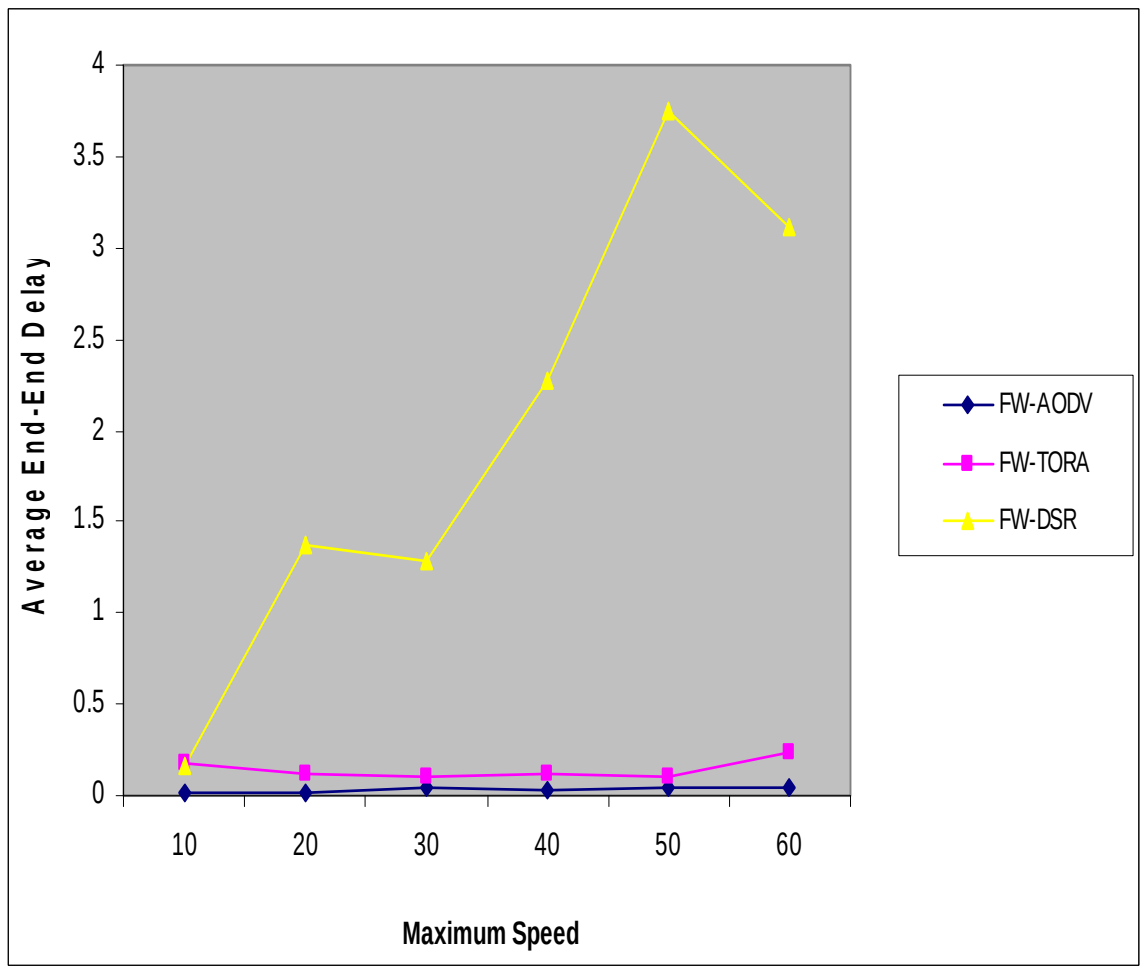

Figure 6(c) Average End-End Delay for AODV, TORA and DSR on Freeway Mobility Model

In Figure 6(a), 6(b) and 6(c) we observe that DSR once again fairs poorly for the metrics. This is mainly because DSR was meant for small diameter topologies and the problem is further compounded by very low link duration and spatial dependency metric for Freeway Mobility Model. AODV once again performs well relatively to TORA and DSR routing protocols. The low degree of spatial dependency and high relative speed in Freeway

Mobility Model also affects the performance of TORA and DSR and to some extent AODV as compared to our proposed model GVMM. Thus AODV can be considered for further studies in applications where nodes mimic vehicular movement

## 5 Conclusion

In this paper we have visualized the application of Mobile Ad-Hoc Routing protocols like TORA, AODV and DSR for nodes which move in a vehicular type of fashion along fixed pathways and lanes. We have considered Manhattan and Freeway mobility models. We have also proposed a Group Mobility model known as Group Vehicular Mobility model. Our performance metrics indicate that Group Vehicular Mobility model outperforms other mobility model by high spatial dependency, high link duration and low relative speed. Thus from this analysis we observe that the routing exhibit higher packet delivery fraction over GVMM. Also the low spatial dependency, link duration and high relative speed affects the performance of routing protocols under Manhattan and Freeway mobility models. Thus characterization of group mobility and realization in terms of vehicular movement with geographic restrictions has aided the robustness of routing protocols in ad hoc networks. In future we would like to experiment with the displacement and direction of mobility nodes with traces of vehicles taken from realistic traffic scenarios.

## 6 References

[1] X. Hong, M Gerla, G PEI, C-C Chiang, A Group Mobility Model for Ad Hoc Wireless Networks *Proc of ACM/IEEE MSWiM* , August 1999.

[2] S Kurkowski, T Camp and W Navidi, Minimal Standards for Rigorous MANET Routing Protocol Evaluation, Technical Report MCS 06-02, Colorado School of Mines, 2006.

[3] S.A.Kulkarni, Dr. G. R Rao, "Mobility and Energy Based Performance Analysis of Temporally Ordered Routing Algorithm for Ad hoc Wireless Network", *Special Issue of IETE on Next Generation Network Converging Regime*, vol. 25, Issue 4, pp 222 – 227, July - August 2008.

[4] G Pei, M Gerla, X Hong, C-C Chiang, A Wireless Hierarchical Protocol with Group Mobility in IEEE WCNC, September, 1999.

[5] ETSI, Universal Mobile Telecommunications System (UMTS), "Selection procedures for choice of radio transmission technologies of the UMTS", UMTS 30.03 Version 3.2.0,. Available at http://www.3gpp.org/ftp /specs/html-info/3003u.htm, 1998

[6] Bai F, Sadagopan N, Helmy A. The IMPORTANT framework for analyzing the Impact of Mobility on Performance of RouTing protocols for Adhoc NeTworks. *Elsevier Ad Hoc Networks* vol. 1, pp. 383-403, 2003

[7] NS-2, "The NS-2 Simulator", http://www.isi .edu/nsnam/ns, 2006.

[8] Perkins C and Royer E. Ad Hoc On-Demand Distance Vector Routing, *Proc of the 2nd IEEE Workshop on Mobile Computing Systems and Applications*. pp 90-100, 1999.

[9] Charles Perkins, Elizabeth Royer and Samir Das, "Ad hoc on demand distance vector (AODV) routing", http://www.ietf.org/internet -drafts/draft-ietf-manet-aodv-03.txt, 1999.

[10] Josh Broch, David Johnson and David Maltz, "The dynamic source routing protocol for Mobile ad hoc networks", http://www.ietf .org/internet-drafts/draft-ietf-manet-dsr-0.1.txt, 2000.

[11] David B. Johnson, David A Maltz, Josh Broch, "DSR: The Dynamic Source Routing Protocol for Multi-hop Wireless Ad Hoc Networks", http://citeseer.nj.nec.com/broch99 supporting. html, 2001.

[12] V D Park and M S Corson, "A Highly Adaptive Distributed Routing Algorithm for Mobile Wireless Networks", *Proc of INFOCOM 97*, April 1997.

[13] J Broch, D A Maltz, D B Johnson, Y C Hu and J Jetcheva, " A performance comparison of multi-hop wireless ad hoc routing protocols", *Proc of the 4th International Conference on Mobile Computing and Networking (ACM MOBICOM 98)*, pp.85-97, October 1998,

**Shrirang.Ambaji.Kulkarni** obtained his Bachelors Degree B.E. from Karnataka Univeristy Dharwar in 2000 and obtained Masters Degree M.Tech. in Computer Network Engineering from Visvesvaraya Technological University Belgaum in the year 2004. Presently he is a Ph.D. research scholar in the Department of Computer Science and Engineering at N.I.E Mysore, India. His area of interest includes routing in ad-hoc networks, reinforcement learning algorithms and energy conservation issues in ad-hoc networks. He is life member of CSI, ISTE

**Dr. G Raghavendra Rao** obtained his Ph.D. from University of Mysore in the year 1999. He is presently working as the Principal of N.I.E Institute of Technology, Mysore. His areas of interests include genetic algorithms, wireless networks cryptography and network security. He is a member of IEEE, ACM and ISTE.